\documentclass[a4paper,11pt]{article}
\pdfoutput=1 

\usepackage{jcappub}
\usepackage{slashed}
\usepackage{graphicx}
\usepackage{dcolumn}
\usepackage{bm}
\usepackage[dvipsnames]{xcolor}

\usepackage[T1]{fontenc} 

\newcommand{\oii}{{[{\sc O\,ii}]}}

\title{Mitigating the impact of fiber assignment on clustering measurements from deep galaxy redshift surveys}

\author[a,1]{Tomomi Sunayama,\note{Corresponding author.}}
\author[a]{Masahiro Takada,}
\author[b]{Martin Reinecke,}
\author[a]{Ryu Makiya,}
\author[c,a]{Takahiro Nishimichi,}
\author[b,a]{Eiichiro Komatsu,}
\author[d]{Shun Saito,}
\author[a]{Naoyuki Tamura,}
\author[a]{Kiyoto Yabe}

\affiliation[a]{Kavli Institute for the Physics and Mathematics of the Universe (WPI), The University of Tokyo Institutes for Advanced Study (UTIAS),
The University of Tokyo, 5-1-5 Kashiwanoha, Kashiwa-shi, Chiba, 277-8583, Japan}
\affiliation[b]{Max-Planck-Institut f\"ur Astrophysik, Karl-Schwarzschild-Str. 1, 85748 Garching, Germany}
\affiliation[c]{Center for Gravitational Physics, Yukawa Institute for Theoretical Physics, Kyoto University, Kyoto, 606-8502, Japan}
\affiliation[d]{Department of Physics, Missouri University of Science and Technology, 1315 N. Pine Street, Rolla, MO 65409, U.S.A.}

\emailAdd{tomomi.sunayama@ipmu.edu}

\abstract{We examine the impact of fiber assignment on clustering measurements from fiber-fed spectroscopic galaxy surveys. We identify new effects which were absent in previous, relatively shallow galaxy surveys such as Baryon Oscillation Spectroscopic Survey (BOSS). Specifically, we consider deep surveys covering a wide redshift range such as $0.6\le z\le 2.4$, as in the Subaru Prime Focus Spectrograph (PFS) survey. Such surveys will have more target galaxies than we can place fibers on. This leads to two effects. First, it eliminates fluctuations with wavelengths longer than the size of the field of view, as the number of observed galaxies per field is nearly fixed to the number of available fibers. We find that we can recover the long-wavelength fluctuation by weighting galaxies in each field by the number of target galaxies. Second, it makes the preferential selection of galaxies in under-dense regions. We mitigate this effect by weighting galaxies using the so-called ``individual inverse probability''. Correcting these two effects, we recover the underlying correlation function at better than $1\%$ accuracy on scales greater than $10~h^{-1}~{\rm Mpc}$.}

\begin{document}

\begin{flushright}
        \quad \\
        \quad \\
        \quad \\
        YITP-19-121\\
\end{flushright}

\maketitle
\flushbottom

\section{Introduction}
\label{sec:intro} 
Clustering of galaxies can be faithfully measured from spectroscopic surveys as long as the selection of galaxies is fair. However, practical constraints often violate fair sampling. 

The well known example is the fiber collision. Fibers cannot be placed to galaxies which are too close to each other in the angular separation. This alters the measured clustering because spectra of targets in over-dense regions are taken less frequently. There are two approaches to mitigate this effect. One approach is the so-called ``nearest neighbor'' method \cite{Zehavi_etal2005,Hahn_etal2017,Berlind_etal2006,Yang_etal2019}, which is based on the assumption that two objects close in angular positions are physically close. We then weight galaxies by the number of nearby galaxies we cannot observe. The other approach is to weight galaxy pairs by the angular correlation function of the target galaxies \cite{Hawkins_etal2003,Li_etal2006,White_etal2011,Okumura_etal2016}.

These methods work only for relatively shallow surveys such as SDSS/BOSS \cite{Alam_etal2015}, for which objects close in angular positions are likely physically close, and the angular correlation of target galaxies contains meaningful information. Unfortunately, both conditions are violated by the upcoming deep spectroscopic surveys with the Subaru Prime Focus Spectrograph (PFS) \cite{Takada_etal2014} and the Dark Energy Spectroscopic Instrument (DESI) \cite{desi1,desi2}.
    In the PFS cosmology survey, we observe emission line galaxies (ELGs) in the wide redshift range from $z=0.6$ to $2.4$ over $1400~{\rm deg}^2$. The DESI survey also targets ELGs from $z=0.6$ to $1.6$ over $15000~{\rm deg}^2$. Both PFS and DESI surveys are much deeper than  SDSS/BOSS; thus, it is less likely than BOSS that two galaxies close in the angular separation are physically close. Moreover, the distribution of target galaxies in sky is close to uniform due to a wide redshift range of the surveys. 
    The DESI team considered two ways to unbias clustering measurements. One way is to generate random catalogs with 
    angular positions selected from the observed galaxies \cite{Burden_etal2017,Pinol_etal2017}. The other way is to up-weight galaxy pairs by the inverse of the probability that a pair of galaxies is observed \cite{Bianchi_etal2018,BianchiPercival2017,Smith_etal2019}. 
    
    The PFS survey is significantly deeper than DESI; thus, the above effects are exacerbated. The PFS will have many more target galaxies than it can place fibers on. This leads to a new effect, which we investigate in this paper. We shall show how the PFS cosmology fiber assignment affects clustering measurements, and how to mitigate the issues. The rest of this paper is organized as follows. In Section 2, we give an overview of the PFS cosmology survey, its fiber assignment algorithm, and the galaxy mocks we use for our analysis. In Section 3, we present how the PFS fiber assignment alters the clustering measurement. In Section 4, we discuss how to mitigate the effects. We conclude in Section 5.

\section{Specifications of the PFS cosmology survey}
\label{sec:overview}

The PFS is a massively multiplexed, optical and near-infrared spectrometer, mounted at the prime focus of the 8.2m Subaru Telescope 
\cite{2016SPIE.9908E..1MT}. It is now under construction. Its focal plane is equipped with 2394 robotically reconfigurable fibers distributed over the 1.3-degree wide hexagonal field-of-view, as illustrated in the top panel of Fig.~\ref{fig:tileratio_ets1}\footnote{Also see \url{https://pfs.ipmu.jp/research/parameters.html}}. 
The spectrograph system is designed to cover a wide range of wavelengths from 380 to 1260~nm in a single exposure.

In this section, we provide an overview of the PFS cosmology survey design and discuss details of the fiber assignment algorithm used in the PFS cosmology. We also describe the galaxy mock catalog we use in this paper. 

\subsection{Survey design: tiling and fiber assignment}
The PFS cosmology program aims at mapping the three-dimensional distribution of about 4 million \oii\ 
ELGs up to $z=2.4$ given the wavelength coverage of PFS \cite{Takada_etal2014}. We use the pre-existing, deep multi-color imaging catalog of the Subaru Hyper Suprime-Cam (HSC) survey \cite{2018PASJ...70S...4A} to construct a catalog of target galaxies. The HSC survey has a depth of $i\sim 26$ ($5\sigma$ for a point source with 
$2^{\prime\prime}$ aperture), and 5 passbands $grizy$ in each field over about 1400~deg$^2$. We select target galaxies using magnitude and color information \cite{Takada_etal2014}.

\begin{figure}
\centering
  \includegraphics[width=0.8\textwidth]{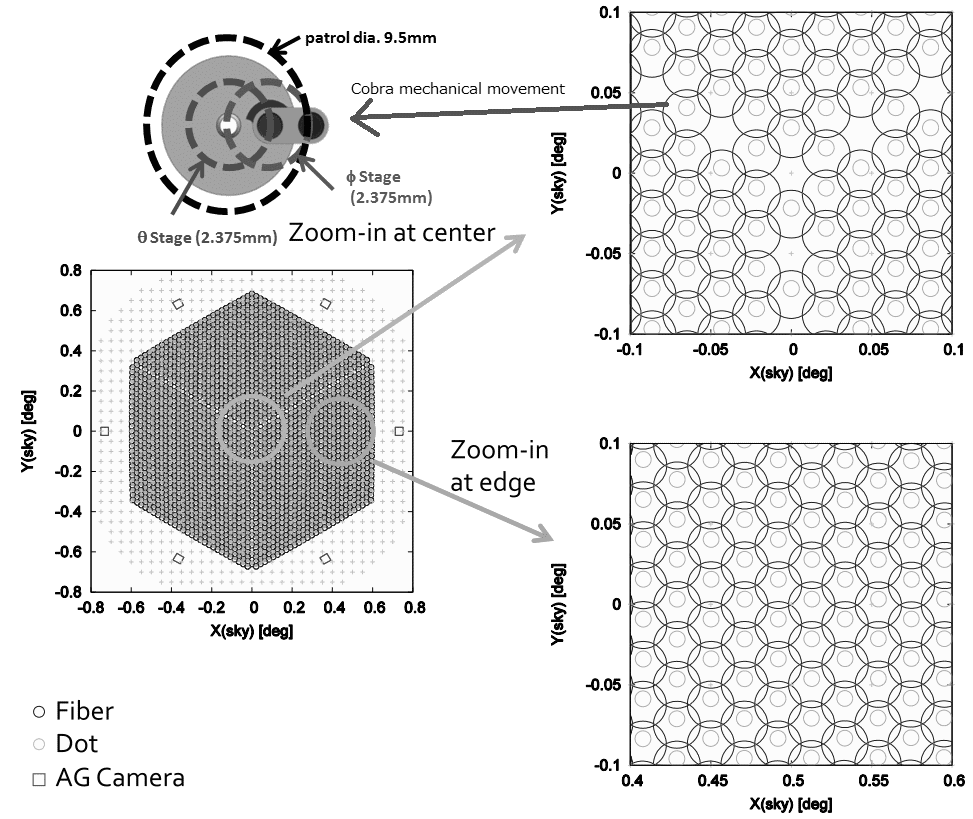}
  \includegraphics[width=0.5\textwidth]{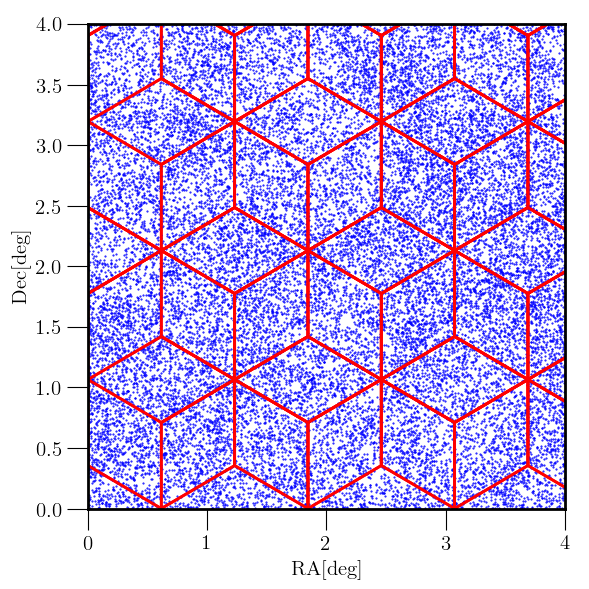}
    \caption{\label{fig:tileratio_ets1} {\it Top}: An illustration of the configuration of 2394 fiber-fed positioners in the focal plane of PFS, taken from Ref.~\cite{Shimono2016}. The shaded hexagon region in the middle left panel denotes the field-of-view corresponding to about 1.25~deg$^2$ on the sky. Each positioner has two motors to cover its patrol area to allocate its fiber to a desired target (denoted by the dashed black circle in the top left panel). The two panels in the right show configurations of the patrol areas of positioners (black circles) and dead ``dot'' regions (grey circles) at two different locations in the focal plane. Target galaxies cannot be observed in the dot regions. {\it Bottom}: The default tiling strategy of the PFS cosmology survey. The blue points show the distribution of target galaxies in $0.6<z<1.2$, taken from the mock catalog.}
\end{figure}

There are two factors to consider, ``completeness'' and ``success rate'', when defining an optimal design/strategy of the PFS cosmology survey. The completeness is defined as the ratio of the number of fiber-assigned (observed) galaxies to the number of targets in each field, whereas the success rate is defined as the ratio of the number of successful redshift measurements to the number of fiber-assigned galaxies.
An optimal survey design will have both
high completeness and high success rate.

To explore an optimal survey design of the PFS cosmology program, we first need to understand the key difference between the PFS and the previous spectroscopic surveys such as BOSS \cite{eisenstein2001}. First, each fiber positioner of PFS has a finite ``patrol'' area within which the fiber can move to a desired target (see Fig.~\ref{fig:tileratio_ets1}). The efficiency of allocating the 2394 fibers to input targets can be calculated as follows. Suppose that each positioner has ``$m$'' target galaxies within its patrol area on average. Since the PFS program observes galaxies over a wide redshift range of $0.6\le z\le 2.4$, the angular distribution of target galaxies on the sky is safely considered as homogeneous; thus, we can assume the number of target galaxies for each positioner follows 
a Poisson distribution with the mean ``$m$'' (here we ignore the overlapping regions of neighboring positioners' patrol areas for simplicity). Assuming a two-visit observation of each field \citep{Takada_etal2014} the efficiency of fiber allocation to targets is given by
\begin{align}
E&= 1- P(n=0) - \frac{1}{2}P(n=1) \nonumber\\
&= 1 - e^{-m} - \frac{1}{2}m e^{-m}, 
\label{eq:efficiency}
\end{align} 
where $P(n)=(m^n/n!)e^{-m}$ is the probability that a fiber (positioner) has $n$ target galaxies in its patrol area. The term with $P(n=0)$ in the first line on the r.h.s. gives a probability that the fiber does not have any target ($n=0$), therefore zero efficiency. The term with $P(n=1)$ gives a probability that the fiber has only one target galaxy; therefore the galaxy is surely observed in either of the two visits, leading to 50\% ($=1/2$) efficiency. 

As shown in Ref.~\cite{Takada_etal2014}, if we feed 8000 target galaxies into each field ($m=8000/2394\simeq 3.3$), the efficiency is $E\simeq 0.90$. With this number of targets, we could make a reasonably good use of the high PFS multiplexity. 
In this case, the completeness is given by
$2\times 2394\times 0.9/8000\simeq 0.54$, i.e.{} 54\% completeness. This can be compared to the BOSS, which reached about 90\% completeness. If we had more than 8000 targets, the completeness would become even lower. 

The low completeness results in non-uniform sampling of galaxies. If a positioner has 1 or 2 targets within its patrol area, those galaxies are surely observed during the two visits. If a fiber has 3 or more targets, only 2 galaxies are observed. This causes a non-uniform (non-random) sampling of target galaxies. Compared to this,  BOSS observed almost all target galaxies due to the high completeness. In addition, most of galaxies which are close in angular separations in the PFS survey are likely due to a chance projection of multiple galaxies at different redshifts. In BOSS, they are likely physically close.

The success rate will multiply the completeness, 
further reducing the number of successful redshift measurements in each field.  
Ref.~\cite{Takada_etal2014} estimated about 74\% success rate using the COSMOS mock catalog \cite{Jouvel2009} that gives a model estimate of \oii\ strength as a function of the HSC broadband photometry of each target galaxy.

With this survey design, we can cover 1400~deg$^2$ with reasonably high number density of galaxies over $0.6\le z\le 2.4$, using about 100 Subaru nights (including the 0.7 weather factor). The lower panel of Fig.~\ref{fig:tileratio_ets1} illustrates the survey strategy; it has two visits for each field with a large dithering pattern, separated by about radius of the hexagonal Field-of-View (FoV) between the two visits. Throughout this paper, we call each hexagonal region in the lower panel of Fig.~\ref{fig:tileratio_ets1} a ``tile''. 

In summary, the PFS cosmology survey differs from BOSS in three aspects:
\begin{itemize}
\item The PFS survey has a low ($\approx$ 50\%) completeness, whereas BOSS has about 90\% completeness.
\item In the PFS survey, we have 4788 available fibers for roughly 8000 target galaxies in each field with two visits. In  BOSS, there are 1000 available fibers for roughly 600 galaxies, which enables almost uniform sampling.
\item The non-uniform sampling of target galaxies due to the limited patrol area of each fiber positioner.  
\end{itemize}
These differences cause different systematic effects in the clustering measurements from the PFS survey, and we develop a method to correct them.

\subsection{Exposure Targeting Software (ETS)}
\label{sec:ETS}
To simulate the fiber assignments to target galaxies in a galaxy survey, we use the fiber assignment software, Exposure Targeting Software (ETS)\footnote{The code is publicly available at \url{https://github.com/Subaru-PFS/ets_fiber_assigner}.}, developed by the PFS project \cite{Shimono2016}.
The ETS requires an input file containing RA and DEC of target galaxies with priority. We run the ETS on the mock catalogs to simulate the fiber assignment, which allows us to include all possible effects of tiling and fiber assignments on the clustering measurements. Here we assume the same priority for all mock target galaxies.

We have several options to choose for an algorithm to determine how to assign 
each fiber to a target in the list. In this paper we employ a ``naive'' algorithm, which assigns a fiber to a target with the highest priority among all the available targets within the patrol area. If there are multiple targets with the same priority, it selects one target randomly.

\subsection{Light-cone mock catalogs of target galaxies in a PFS-like survey}
\label{sec:mocks}
To make a quantitative study of the effects of tiling and fiber assignment, we use mock catalogs of target galaxies in a light-cone volume of $0.6\le z\le 2.4$. We generate the mock catalogs using the publicly available code, 
${\tt lognormal\_galaxies}$\footnote{\url{http://wwwmpa.mpa-garching.mpg.de/~komatsu/codes.html}.},  
as developed in Ref.~\cite{Agrawal_etal2017}.

The code generates the grid-based log-normal density field with a given input power spectrum. We use the linear matter power spectrum to generate the matter density field (used to compute the velocity field) and the matter power spectrum times the linear bias squared to generate the galaxy density field. The matter power spectrum is calculated by {\tt CAMB}\cite{camb} assuming the $\Lambda$CDM model with the {\it Planck} 2015 cosmological parameters \cite{Planck2015cosmoparams}. The code then generates an integer number of galaxies in each grid from the galaxy density field by the Poisson distribution with mean given by the input. Finally, the code places the generated galaxies at random positions within each grid; that is, we ignore clustering of galaxies at scales below the grid size.

The code also allows us to generate the grid-based peculiar velocity field of galaxies from the underlying matter density field.
First we generate the peculiar velocity field in Fourier space by using the linear continuity equation
\begin{align}
\bm{v}(\bm{k}) = i\mathcal{H}f\frac{\bm{k}}{k^2}\delta_m(\bm{k}),
\end{align}
where $\bm{k}$ is a wavenumber vector of each Fourier grid, $\mathcal{H} \equiv aH$ with the scale factor $a$ and the Hubble expansion rate $H=\dot a/a$, $f = d\ln{D}/d \ln{a}$ is the logarithmic growth rate with the linear growth factor $D$, and $\delta_m$ is the matter density field. We then Fourier transform $\bm{v}(\bm{k})$ to configuration space.
Note that we assume that all the galaxies in the same grid have the same peculiar velocity.

As the log-normal method is an approximated method, it cannot fully reproduce detailed properties of clustering in the large-scale structure such as  higher-order moments. However, it is sufficient for the purpose of this paper, which is to quantitatively evaluate the fiber assignment effects on two-point correlation functions.
An advantage of the log-normal model is that it allows us to generate a large number of mock catalogs in the light-cone volume. 

Following the nominal survey design of the PFS cosmology program given in Ref.~\cite{Takada_etal2014}, we generate mock catalogs of galaxies in 7 redshift slices, as summarized in Table~\ref{table:cone}. The table also gives the assumed linear bias parameter of galaxies in each redshift slice. 
In this paper, we focus on one of the spatially-contiguous regions corresponding to the ``Fall equatorial field'' in the HSC-Wide survey footprints as given in Table~5 of Ref.~\cite{2018PASJ...70S...4A}, since the PFS redshift survey is based on a spectroscopic follow-up observation of the HSC galaxies. The area is about 630~deg$^2$, and in the following we use 100 realizations of the mock catalogs to have sufficient statistics. Due to the configuration of the log-normal simulations, the survey area for the mock catalogs is 560~deg$^2$.
The grid size is 
$97.5/1024\simeq 0.095~{\rm degrees}$ on a side, corresponding to about 3--6~$h^{-1}$Mpc over the redshift range of the survey.

\begin{table}
\begin{center}
\begin{tabular}{c|c|c|c|c|c|c}
\hline 
 $z$-slice & $z$-range & $L_{x}$ & $L_{y}$ & $L_{z}$ &  $10^4\bar{n}_{\rm g}$ & $b_g$\\
  & & [$h^{-1}{\rm Mpc}$] &[$h^{-1}{\rm Mpc}$] &[$h^{-1}{\rm Mpc}$] &$[h^3{\rm Mpc}^{-3}]$
  \tabularnewline
\hline 
\hline 
slice1 & 
[0.6,0.8]
& 2129 & 243 & 401 &   5.4 & 1.18
\tabularnewline
\hline 
slice2 & 
[0.8,1.0]
& 2591 & 296 & 355 & 17.2 & 1.26\tabularnewline
\hline 
slice3 & 
[1.0,1.2]
& 3000 & 343 & 316 &  16.6 & 1.34\tabularnewline
\hline 
slice4 & 
[1.2,1.4]
& 3365 & 385 & 282 & 22.4 & 1.42\tabularnewline
\hline 
slice5 & 
[1.4,1.6]
& 3692 & 422 & 253 &  15.8 & 1.50\tabularnewline
\hline 
slice6 & 
[1.6,2.0]
& 4123 & 471 & 434 & 7.2 & 1.62\tabularnewline
\hline 
slice7 & 
[2.0,2.4]
& 4607 & 527 & 361 &  7.8 &1.78\tabularnewline
\hline 
\end{tabular}
\caption{\label{table:cone}
Details of target galaxies in 7 redshift slices, generated from the log-normal mock catalogs of light-cone volume. 
$\bar{n}_{\rm g}$ denotes the mean number density of target galaxies (not fiber-assigned galaxies) in each redshift slice, and $b_{\rm g}$ is the linear galaxy bias parameter. 
 $L_x$, $L_y$ and $L_z$ denote the comoving side lengths of three-dimensional volume at each redshift, where $L_z$ is the width in the line-of-sight direction. 
}
\end{center}
\end{table}

\section{Fiber assignment artifacts}
\label{sec:ets_effects}
To investigate the effects of fiber assignment on the clustering measurements, we 
run the ETS on each of 100 log-normal mock galaxy catalogs. 
Here we consider the distribution of pointings of the hexagonal FoV as given in the lower panel of Fig.~\ref{fig:tileratio_ets1}. We assume a 100\% success rate of redshift determination for
 fiber-assigned galaxies in the mock; that is, we do not include a failure of the redshift measurement for simplicity as the purpose of this paper is to study the effects of tiling and fiber assignments. 

For each mock
we estimate the two-point correlation function using the 
Landy-Szalay estimator \cite{LandySzalay1993}:
\begin{align}
\label{eq:CF}
\xi(r,\mu)=\frac{DD(r,\mu)-2DR(r,\mu)+RR(r,\mu)}{RR(r,\mu)}, 
\end{align}
where $DD(r,\mu)$ is the number of galaxy pairs separated by $r$ with cosine angle $\mu$ between the direction connecting the pair and the line-of-sight (LOS) direction, $RR(r,\mu)$ is the number of pairs in the random catalog, and 
$DR(r,\mu)$ is the number of pairs of galaxy and random catalogs. Note that these pair counts in bins are normalized appropriately according to the total number of pairs in each category. We use the plane-parallel approximation and take the $z$-axis as the LOS direction.
We compute RA and DEC from comoving coordinates in the transverse directions ($x$ and $y$ coordinates) of galaxies in the mock scaled by the angular diameter distances, such that the side lengths $L_x$ and $L_y$ of the simulation box at each redshift slice scale to 70~deg and 8~deg, respectively.
We use linearly spaced bins; 
$\Delta r=2~h^{-1}{\rm Mpc}$ bin width from $r=2$ to $200~h^{-1}{\rm Mpc}$ and $\Delta \mu=0.05$ in 
$\mu=[0,1]$ (note that we can always choose the pair-separation vector such that the direction cosine $\mu$ is non-negative).
We first generate 10 random samples with the same number density as the target galaxy catalog, and 
run the ETS in each random sample to encode geometrical artifacts due to tiling.
Then, we combine these 10 random samples to create one random catalog.

\begin{figure*}
\centering
    \includegraphics[width=0.48\textwidth]{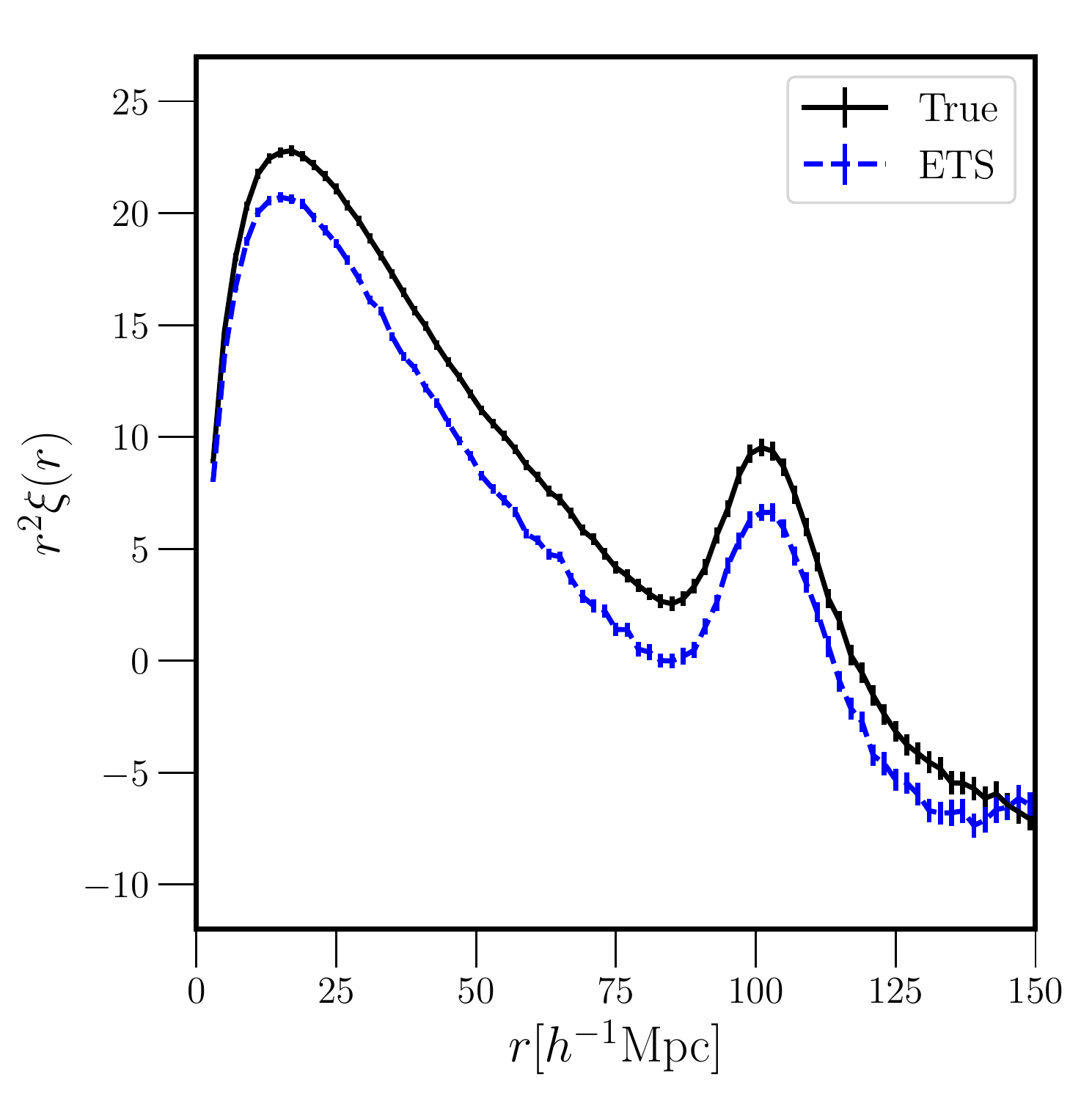}
    \includegraphics[width=0.48\textwidth]{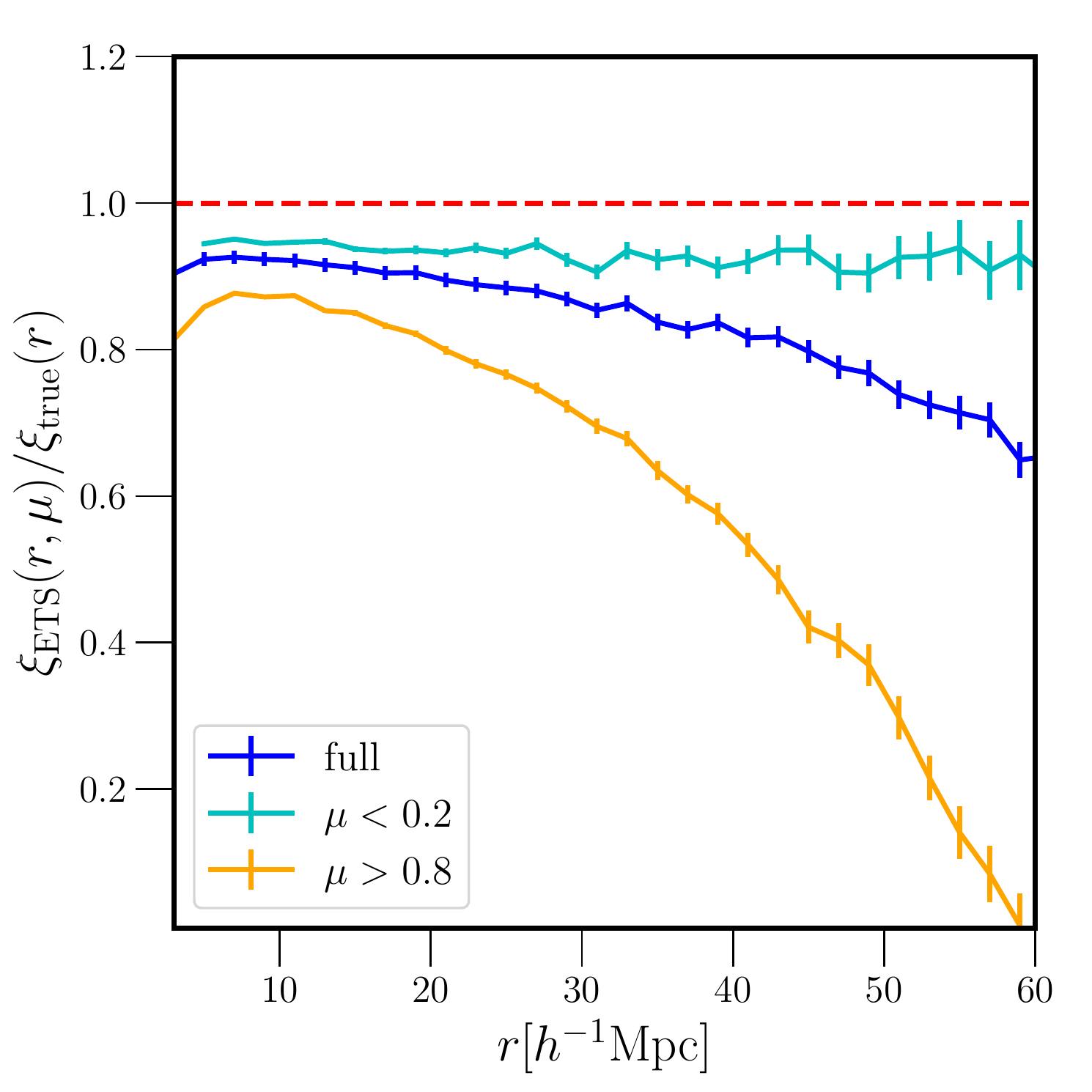} 
    \caption{\label{fig:ets1}
{\it Left}: The blue data with error bars show the real-space correlation function measured from 100 realizations of the mock catalogs with the ETS. Here we show the result from the $z=1.3$ slice. The error bar shows an error on the mean for the survey area of 560~deg$^2$. The data points at different bins are highly correlated. The black data show the underlying correlation function measured from the same mocks. {\it Right}: 
The blue curve shows the ratio of the correlation functions in the left panel.
The cyan and orange curves show the ratios for the correlation functions in 
$\mu<0.2$ and $\mu>0.8$, respectively, where $\mu$ is the cosine angle between the separation and the LOS direction. Note that we only show the ratios up to $r<60~h^{-1}{\rm Mpc}$ because the correlation functions cross zero at larger $r$.
}
\end{figure*}

Here we consider the real-space correlation function for clarity of the nature of problems. 
The left panel of 
Fig.~\ref{fig:ets1} compares the correlation functions of galaxies before and after the ETS at $z=1.3$. 
The redshift slice of $z=1.3$ has the highest number density (see Table~\ref{table:cone}), and the results from the other redshift slices are similar. 
We here consider the correlation function averaged over the angle, i.e., the $\mu$ bins.
We find that the ETS modifies the correlation function at all scales, which may appear surprising at first because the comoving length of one tile at $z=1.3$ is $50~h^{-1}{\rm Mpc}$ in the transverse direction. We explain the reason behind this in the next Section.
These results are different from those of Refs.~\citep{Burden_etal2017, Pinol_etal2017} from a simulation of the DESI-like fiber assignments. They showed a smaller systematic error in the correlation function, and the error exists only at $r<100~h^{-1}{\rm Mpc}$.
However, they considered a single redshift slice (treated a single simulation box for the redshift survey), and did not include multiple redshift slices taken from the light-cone volume. 

The right panel shows the ratio of the correlation functions before and after the ETS. We show the ratios for the angle-averaged correlation functions (blue) as well as for those of transverse ($\mu<0.2$) and LOS ($\mu>0.8$) correlation functions. We find that the fiber assignment produces a $\mu$-dependent correlation function even for the real-space galaxy distribution (i.e. we ignored redshift space distortion in each mock). We find that the LOS direction is more strongly affected by the fiber assignments. Any anisotropy shown here is due to a non-uniform sampling of target galaxies that violates statistical isotropy.

\section{Mitigation of tiling and fiber-assignments effects in the clustering measurements}
\label{sec:random}
The bias seen in Fig.\ref{fig:ets1} is caused by two effects: the tiling of the survey field and the fiber assignment. In this Section, we explain and correct each effect.

\subsection{Tiling}
\label{subsec:tiling}

\begin{figure}
\centering
    \includegraphics[width=0.9\textwidth]{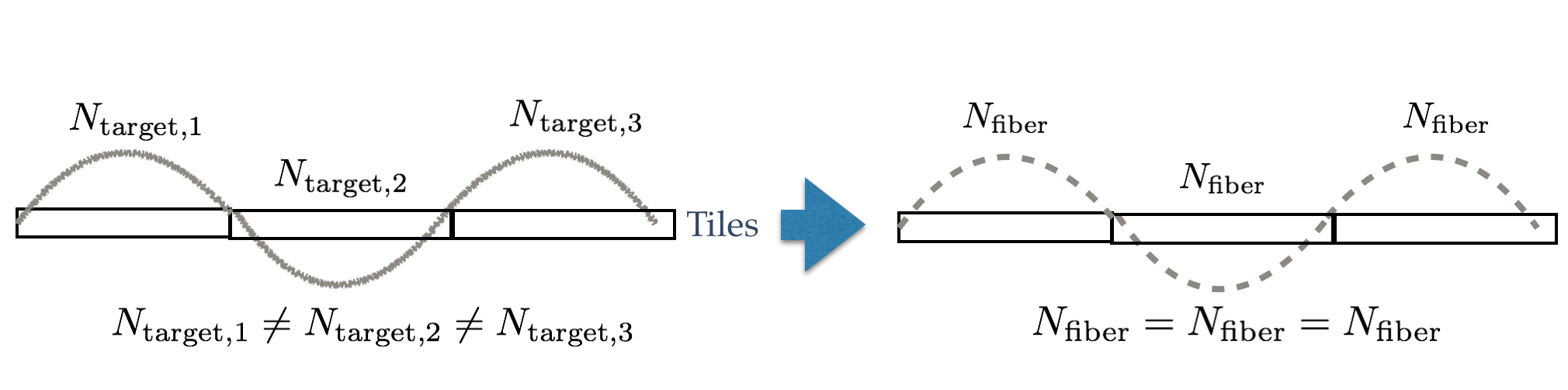}
    \caption{\label{fig:pic_long} 
An illustration of the tiling effect on the clustering measurement for a low-completeness (e.g.{} 50\%) survey such as the PFS survey.
The rectangular shapes represent individual tiles. The number of targets in each tile is generally modulated by the density fluctuations of large-scale structure with wavelengths greater than the tile size (the gray curve in the left figure). If we can observe only a fixed number of galaxies in each tile due to the fixed number of fibers in the focal plane of the spectrograph, we cannot capture the long-wavelength modes. 
    }
\end{figure}

Fig.~\ref{fig:pic_long} illustrates the problem. Limiting the number of observed galaxies to the number of available fibers reduces the correlation function amplitude, when the number of target galaxies is much larger than the number of fibers.
The number of target galaxies varies from tile to tile because of the density fluctuation with wavelengths longer than the tile size.
However, this long-wavelength mode fluctuation in each tile cannot be captured if the number of observed targets is limited to the number of available fibers.
This was not a problem for the previous spectroscopic galaxy surveys such as BOSS, because the completeness 
was quite high, above 90\% as described in Section~\ref{sec:overview}.
On the other hand, the PFS cosmology survey aims for only 50\% completeness.

\begin{figure}
\centering
    \includegraphics[width=0.5\textwidth]{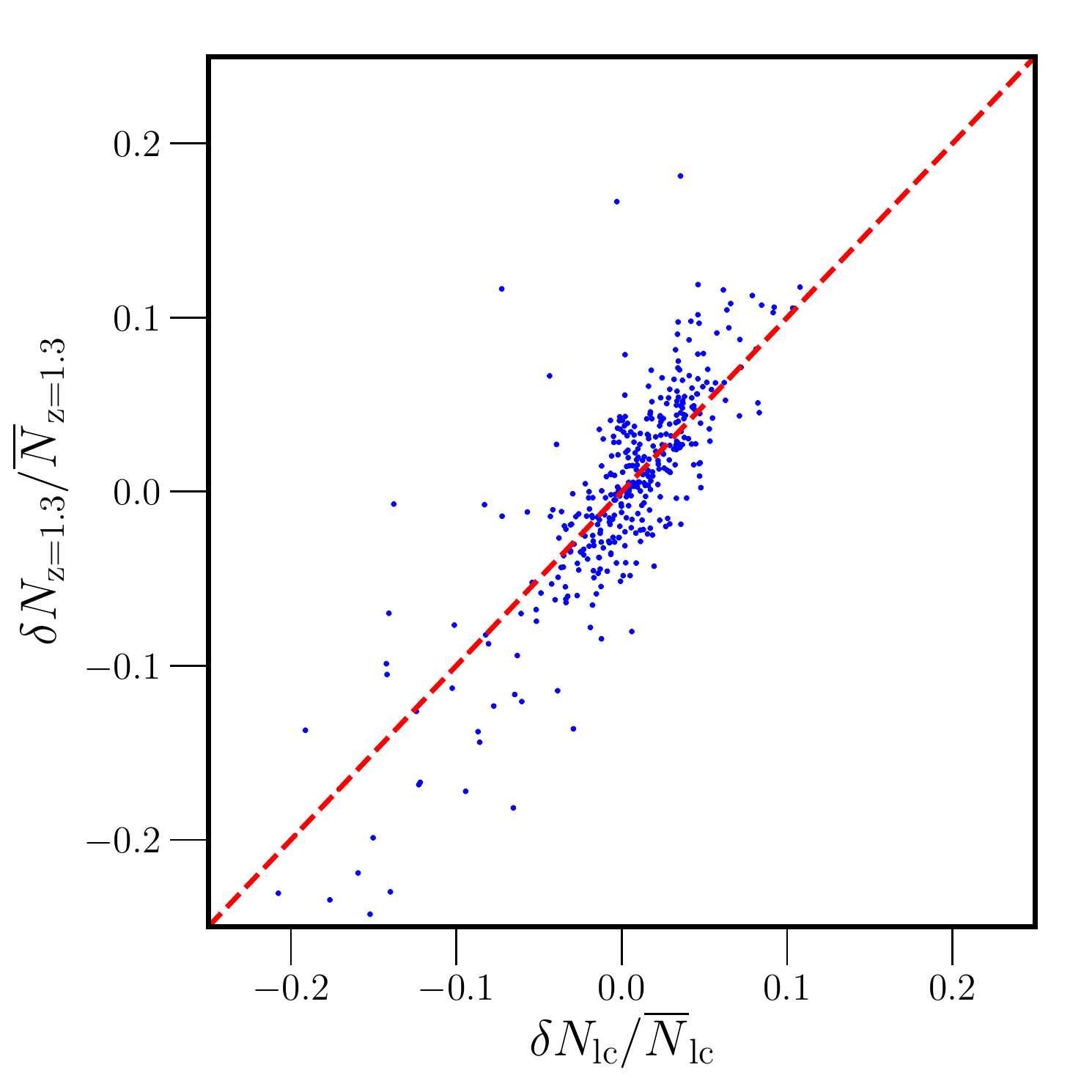}
    \caption{\label{fig:scatter_weight} 
The scatter distribution of the number of targets, measured from one realization of the mock catalog with the ETS.
Each point represents one tile, and there are 377 points corresponding to 560~deg$^2$ (simulated area) in total. 
The $x$-axis denotes 
the number density fluctuation in each tile from the entire light-cone volume,
$\delta N_{\rm lc}/\bar{N}_{\rm lc}$, which is a direct observable in the actual observation. The $y$-axis denotes the the number density fluctuation at one particular redshift slice at $z=1.3$, 
$\delta N_{\rm z=1.3}/\overline{N}_{\rm z=1.3}$, among 7 slices in the light-cone volume.
 }
\end{figure}

To recover the long-wavelength fluctuations, we weight observed galaxies in each tile by
\begin{align}
\label{eq:CF}
w_i=\frac{N_{\rm target,i}}{N_{\rm obs,i}}, 
\end{align}
where $N_{\rm target,i}$ is the number of target galaxies in the $i$-th tile and $N_{\rm obs,i}$ is the number of observed galaxies. 
The modulation in the number of targets over different tiles is a result of the LOS projection, and the modulation does not necessarily reflect the number modulation of targets in a particular redshift slice. 
Fig.~\ref{fig:scatter_weight} shows the correlation between $\delta N_{\rm lc}/\bar{N}_{\rm lc}$ and $\delta N_{\rm z=1.3}/\bar{N}_{\rm z=1.3}$, where $\delta N=N_i-\bar{N}$ and $N_i$ is the number of galaxies in the $i$-th tile.
We find that the modulation in the number of targets over the light-cone volume is strongly correlated with the modulation in a particular redshift slice. We find similar correlations for the other redshift slices.

\begin{figure*}
\centering
    \includegraphics[width=0.48\textwidth]{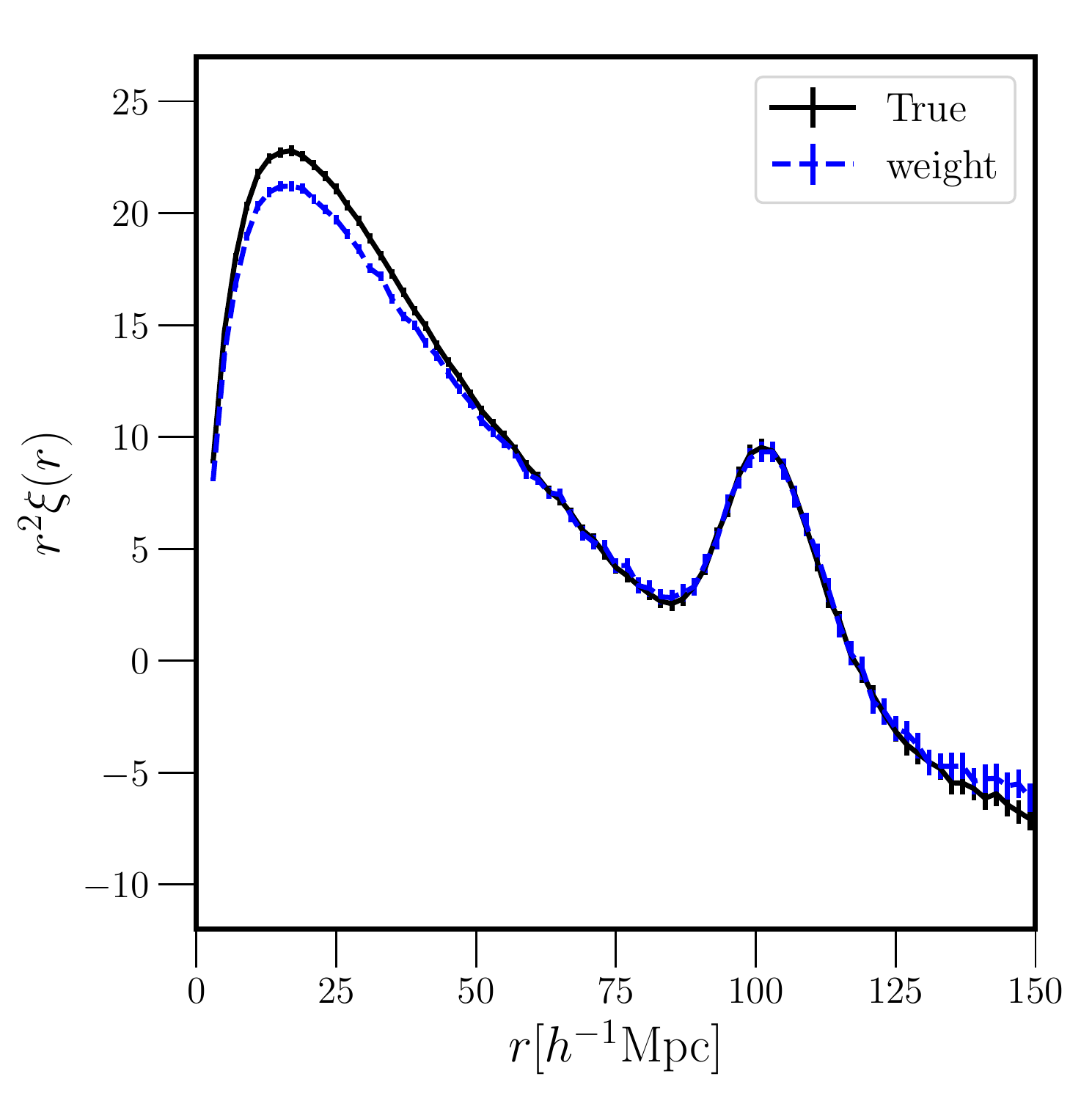} 
    \includegraphics[width=0.48\textwidth]{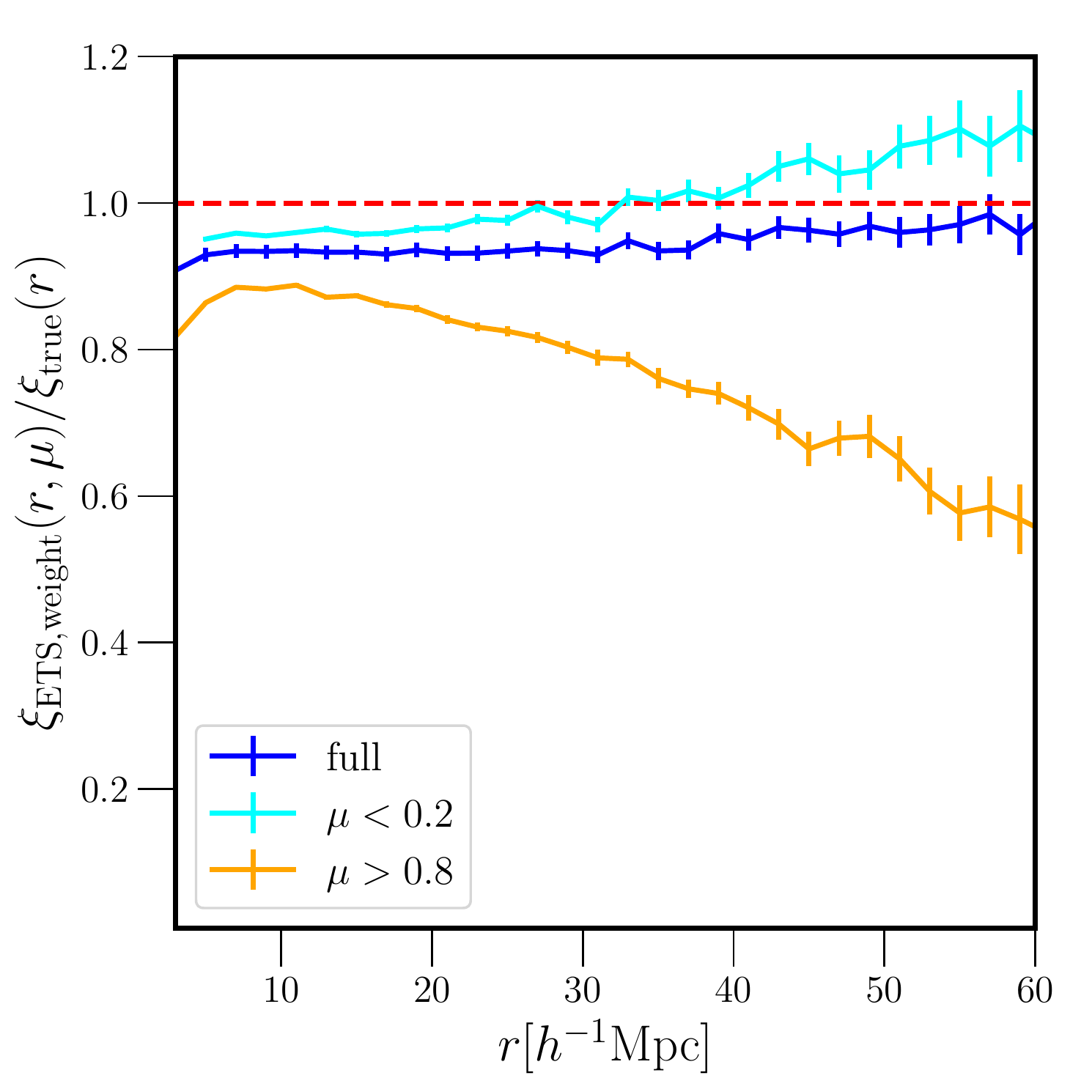}
    \caption{\label{fig:ratio_weight}
    Same as Fig.~\ref{fig:ets1}, but with the weights given by 
    Eq.~(\ref{eq:CF}) applied.
    }
\end{figure*}

Fig.~\ref{fig:ratio_weight} shows the performance of the weighting method defined above.
Compared to Fig.~\ref{fig:ets1}, the weighting scheme recovers clustering on scales $r>60~h^{-1}{\rm Mpc}$ and reduces the scale-dependent deviation on small scales. However, it does not correct all the effect.

\begin{figure*}
\centering
    \includegraphics[width=0.5\textwidth]{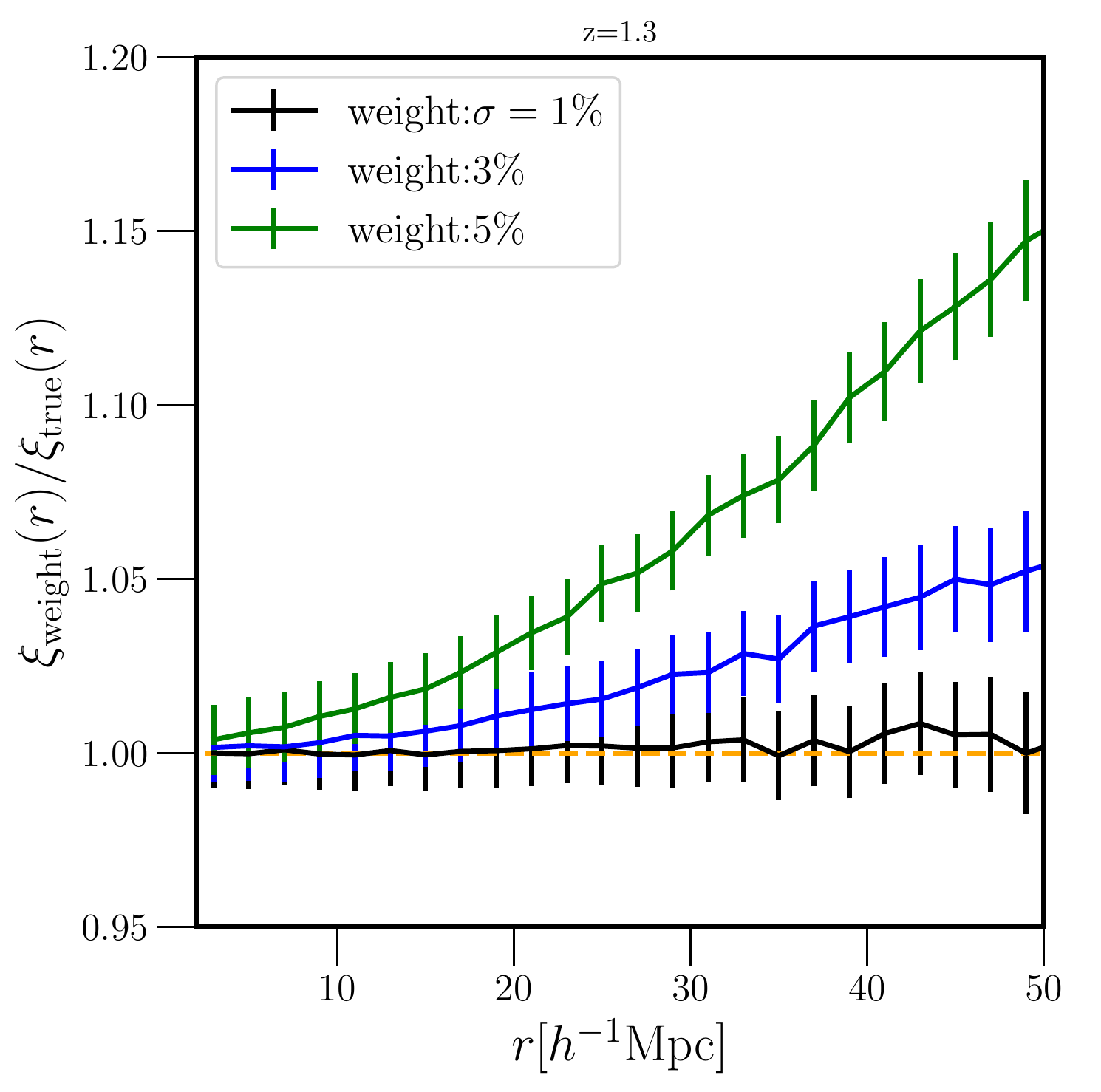}
    \caption{\label{fig:ratio_uncert}
    The ratio of correlation functions  between "true" and after applying the weights given by Eq.~(\ref{eq:CF}) with 1\% (black), 3\% (blue), and 5\% (green) uncertainties. 
    }
\end{figure*}

Before going to the next weighting method to correct for the remaining effect, we comment on required accuracy of the weighting method given by Eq.~(\ref{eq:CF}). We use photometric data to select our targets. However, the photometric error brings galaxies outside the selection into the target list, contaminating the estimated number of desired galaxies. 
Moreover, any masks may lead to uncertainty in the actual number of target galaxies in each tile. In Fig.~\ref{fig:ratio_weight} we assumed no photometry error or no mask. 
To test the impact of these effects, we isolate the effect of tiling by ignoring the fiber assignment. 
Instead of running the ETS, we randomly select target galaxies in each tile with the fixed number of observed galaxies. 
In this way, we will not have any residual bias in correlation functions after applying the correct weights.
We simulate uncertainties in the number of target galaxies by a Gaussian distribution with the true target number as mean. 
Fig.~\ref{fig:ratio_uncert} shows the results, indicating 
that this weighting scheme requires 1\% precision in the target number. 
This puts requirements on the quality of imaging surveys used for target selection.

In principle, the Subaru FMOS galaxy redshift survey (FastSound) \cite{Okumura_etal2016} could have had the same issues. In practice, their low success rate rendered this effect subdominant due to the large statistical error. Therefore, they included the variations of the success rates only in the random catalog and could still recover the underlying correlation functions without weighting observed galaxies.

\subsection{Fiber assignment}
\label{sec:ETS2}
\subsubsection{Toy Model}

\begin{figure*}
\begin{center}
    \includegraphics[width=0.5\textwidth]{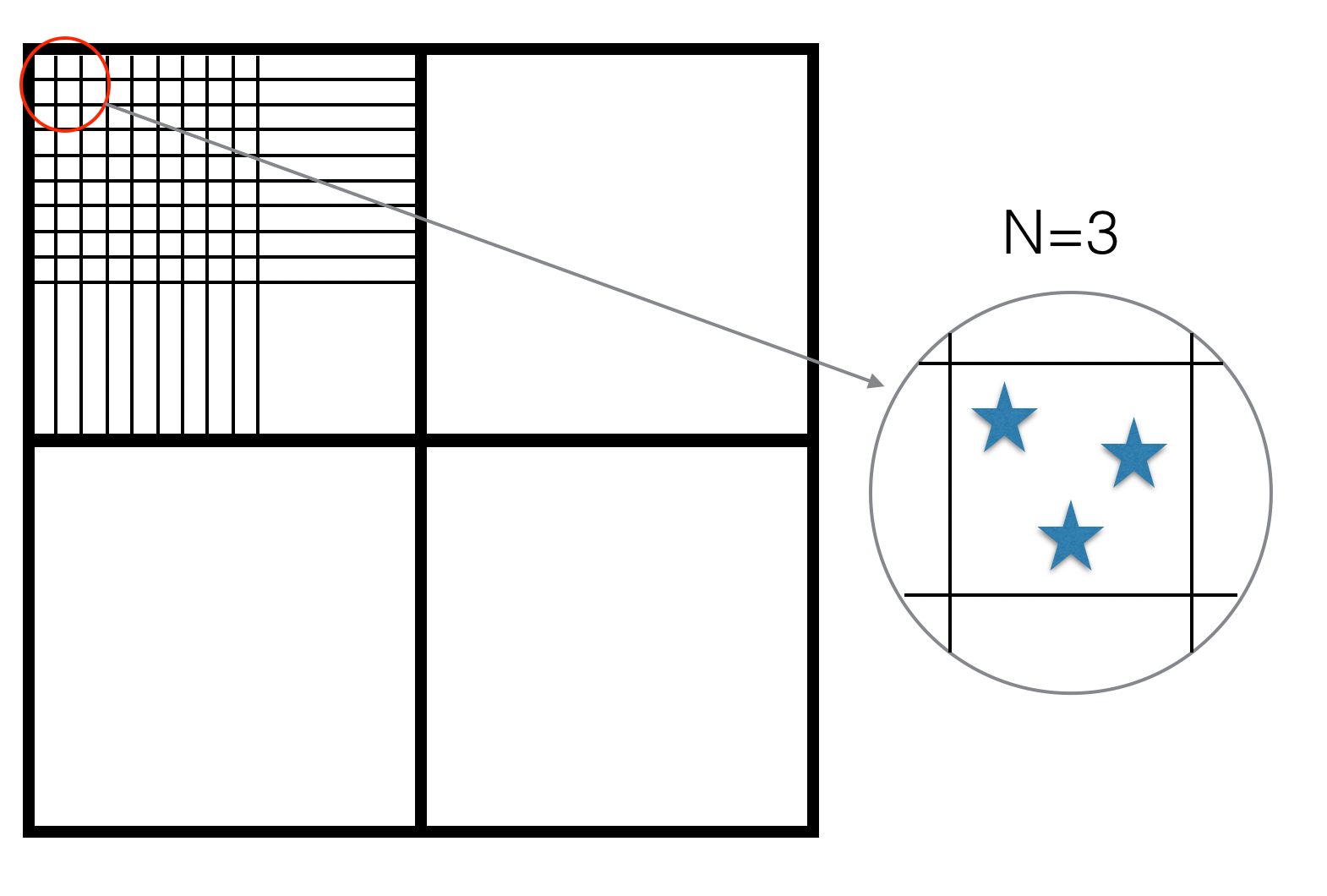}
    \caption{\label{fig:pic2} An illustration of the toy model that we use to mimic the tiling and fiber assignments. 
    Each of the four squares represents one tile, and $50\times 50$ grids inside it represent fibers and their patrol areas.  The zoom-in illustration
    shows one patrol area with three target galaxies.
    }
\end{center}
\end{figure*}
The remaining issue is under-sampling of over-dense regions due to fiber assignment.
To understand the issue, we first use a simplified setup as illustrated in Fig.~\ref{fig:pic2}.
We show four tiles with a side length of 1~degree approximating the PFS FoV and $50\times 50$ grids inside each tile approximating the fibers and their patrol areas in the PFS FoV.
In this setup, we ignore the overlapping patrol areas of neighboring fibers.
We apply these tiling and grid-patrol area configurations to each light-cone realization of the mock galaxy catalogs. 
Assuming two visits for each tile, we randomly select two galaxies from the target list in each grid. 
In this setup,  all the target galaxies  in an under-dense grid (which contains only one or two galaxies) are observed, whereas at most two galaxies are observed in a dense grid. 
This shows the nature of the fiber assignment problem: fibers are assigned to targets in non-uniform manner.

We use a weighting method to correct this problem. 
Specifically, we up-weight galaxies by the inverse of the probability that a target galaxy is observed.
If a fiber has one or two targets in its patrol area,  
they are always observed during the two visits: 100\% completeness. 
If a fiber has three or more targets in its patrol area, each of them has a lower probability to be observed. 
This probability is $2/n_{i,{\rm target}}$ for $n_{i,{\rm target}}\ge 3$, where $n_{i,{\rm target}}$ is the number of targets for the $i$-th fiber.
We then up-weight the fiber-assigned galaxies by $n_{i,{\rm target}}/2$ for $n_{i,{\rm target}}\ge 3$. 
This is called the ``Individual-Inverse Probability'' (IIP) method as proposed in Ref.~\cite{Smith_etal2019}.
This is different from the ``Pairwise-Inverse Probability'' (PIP) method, which up-weights {\it pairs} of galaxies by the inverse of the probability that a galaxy {\it pair} is observed.
However, the PIP method is equivalent to the IIP method in our setup, because we have no fiber collision inside of each grid.

\begin{figure*}
\centering
    \includegraphics[width=0.48\textwidth]{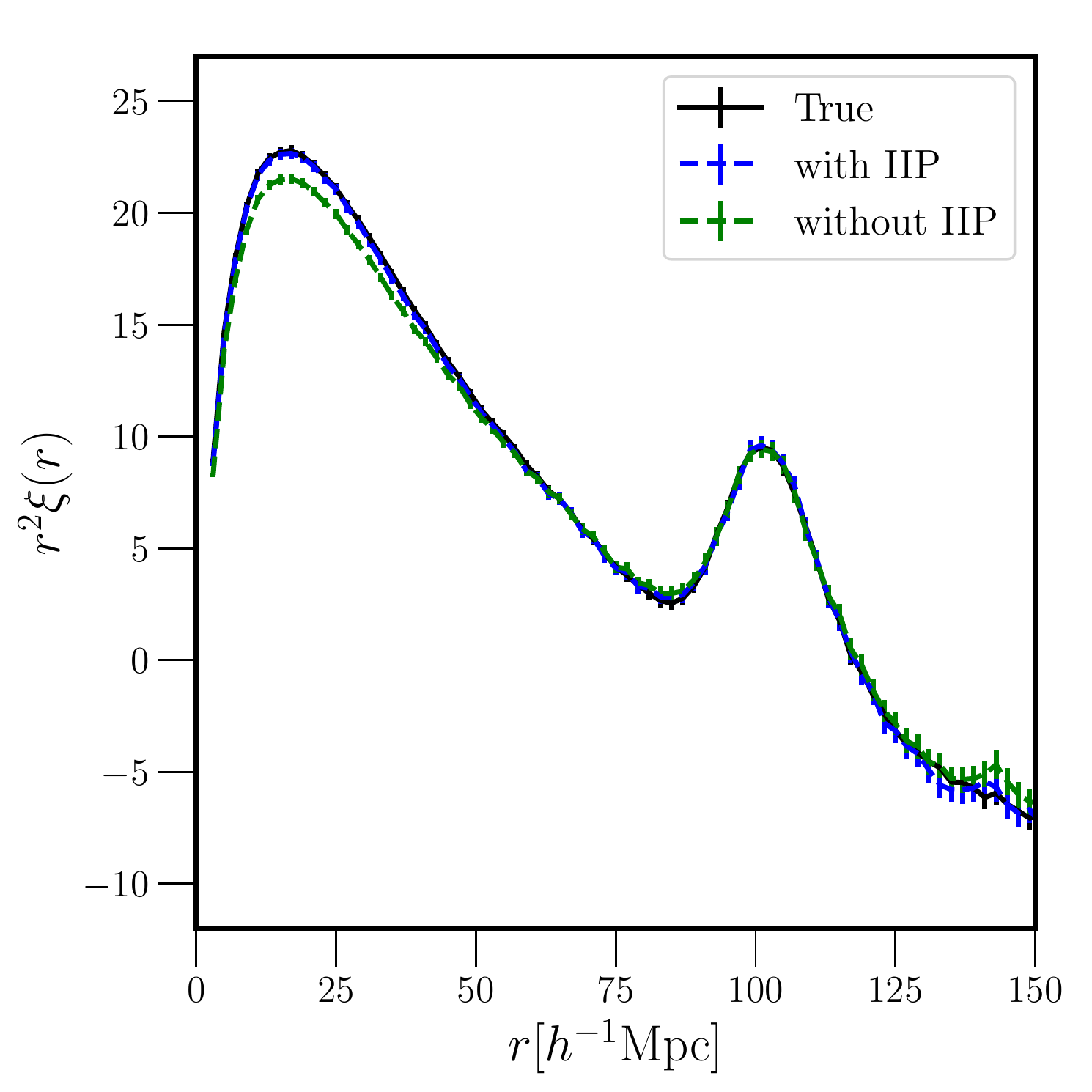}
    \includegraphics[width=0.48\textwidth]{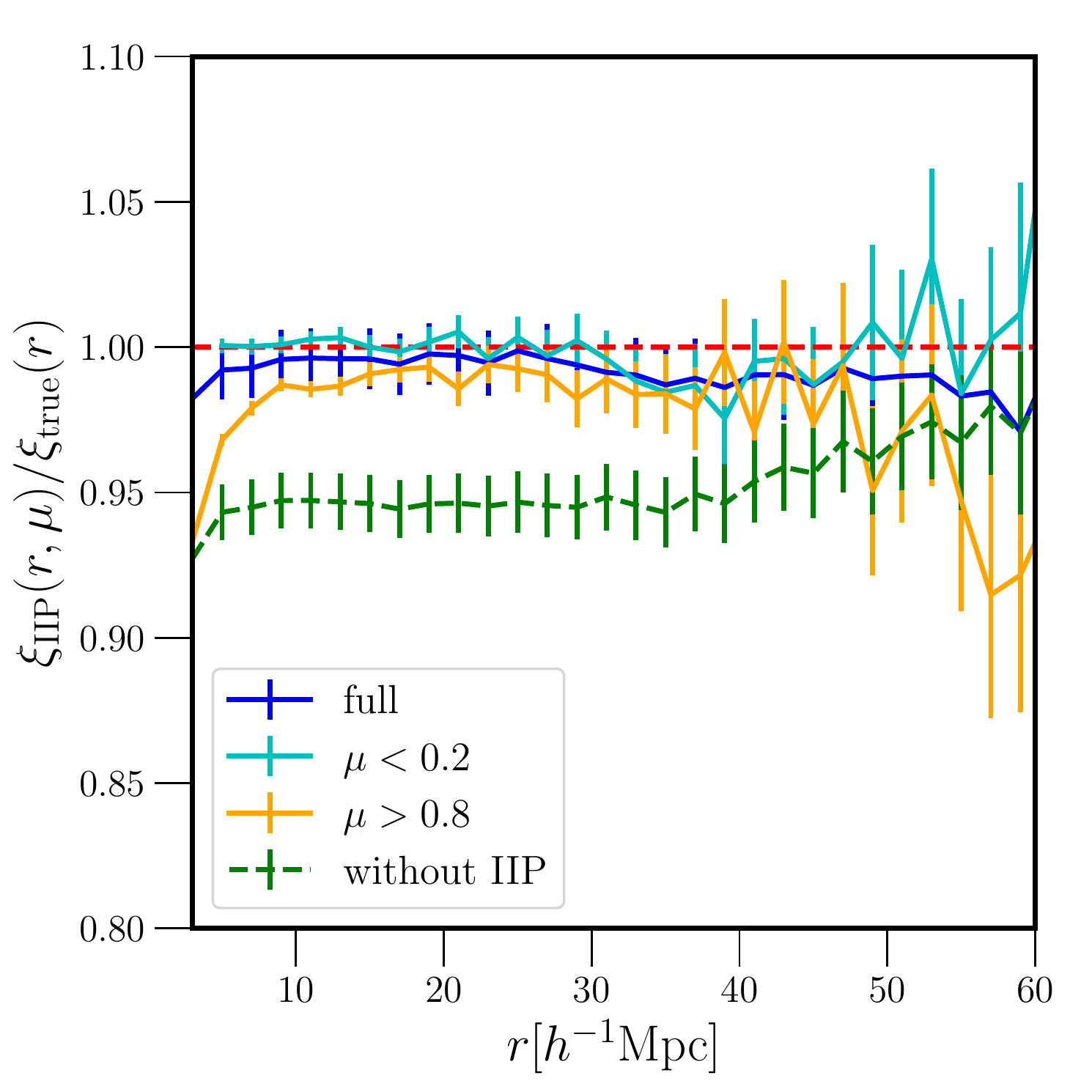}
    \caption{\label{fig:uniform:iip} 
    {\it Left}: The green and blue data with error bars show the real-space correlation functions measured with the fiber assignment illustrated in Fig.~\ref{fig:pic2} before and after applying the IIP, respectively. The black data show the underlying correlation function measured from the same mocks.
    {\it Right}: 
The green and blue curves show the ratios of the correlation functions in the left panel with respect to the underlying correlation function.
The cyan and orange curves show the ratios for the correlation functions in $\mu<0.2$ and $\mu>0.8$, respectively, after applying the IIP.
}
\end{figure*}

In Fig.~\ref{fig:uniform:iip}, we show that this weighting method recovers the underlying correlation functions to better than 1\% accuracy at all separations for the angle-averaged correlation functions. 
The transverse-direction ($\mu<0.2$) is also recovered well, whereas the LOS direction ($\mu>0.8$) is recovered only at $r>10~h^{-1}{\rm Mpc}$.
The correlation functions are not recovered on small scales because the IIP method does not recover the target galaxies we missed, and it only down weights the ones who are preferentially selected. We miss more target galaxies along the LOS in non-random manner due to the number of visits (i.e., 2 visits).

\subsubsection{ETS}
We now apply the IIP method to the mock catalogs with the ETS, which
includes all the relevant effects such as the overlapping patrol areas of neighboring fibers and fiber collisions.

\begin{figure}
\centering
    \includegraphics[width=0.45\textwidth]{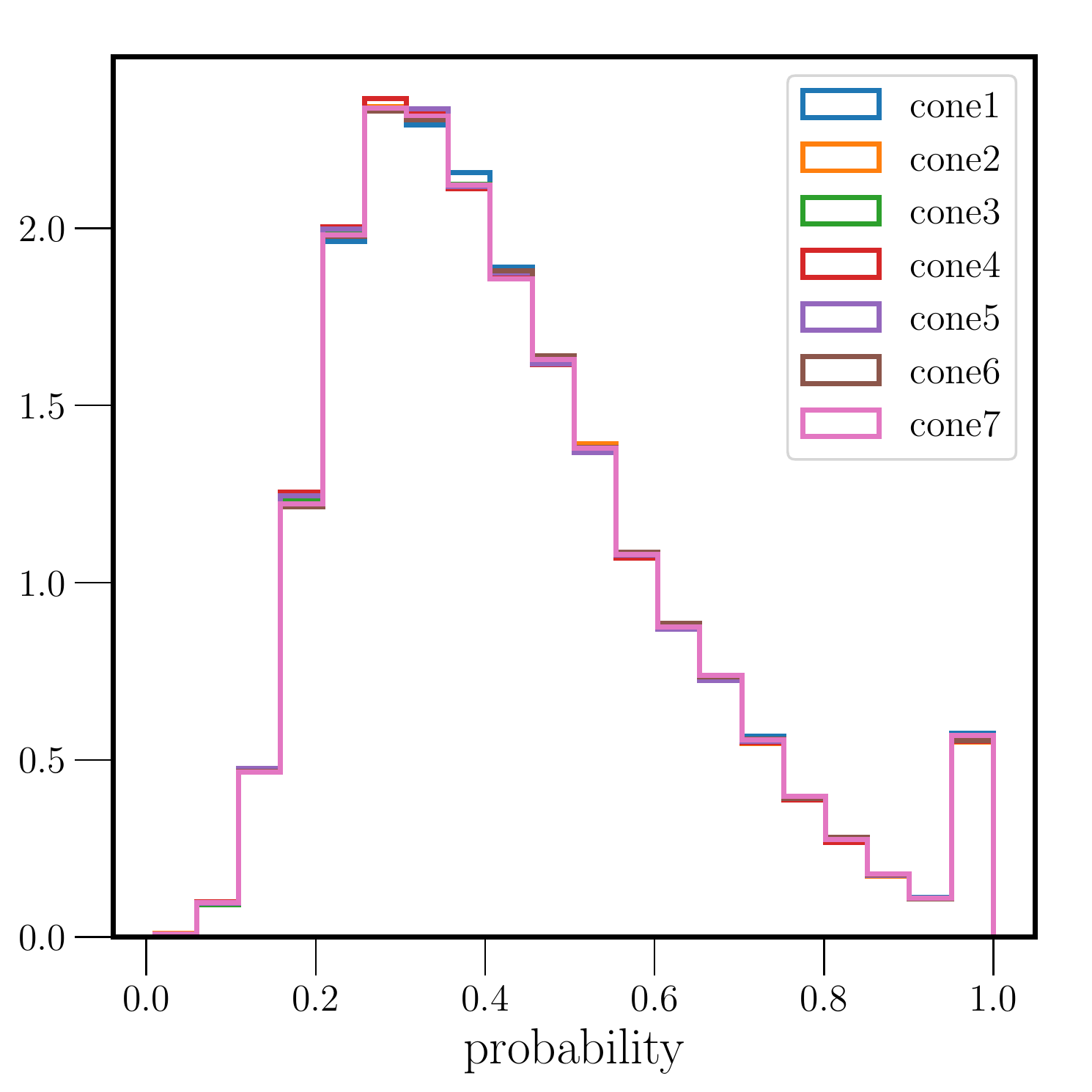} 
    \caption{\label{fig:ets:prob} 
    Probabilities that each target galaxy is observed by fibers
    in the different redshift slices, derived from running the ETS 100 times on one mock catalog.
    }
\end{figure}

\begin{figure*}
\centering
    \includegraphics[width=0.48\textwidth]{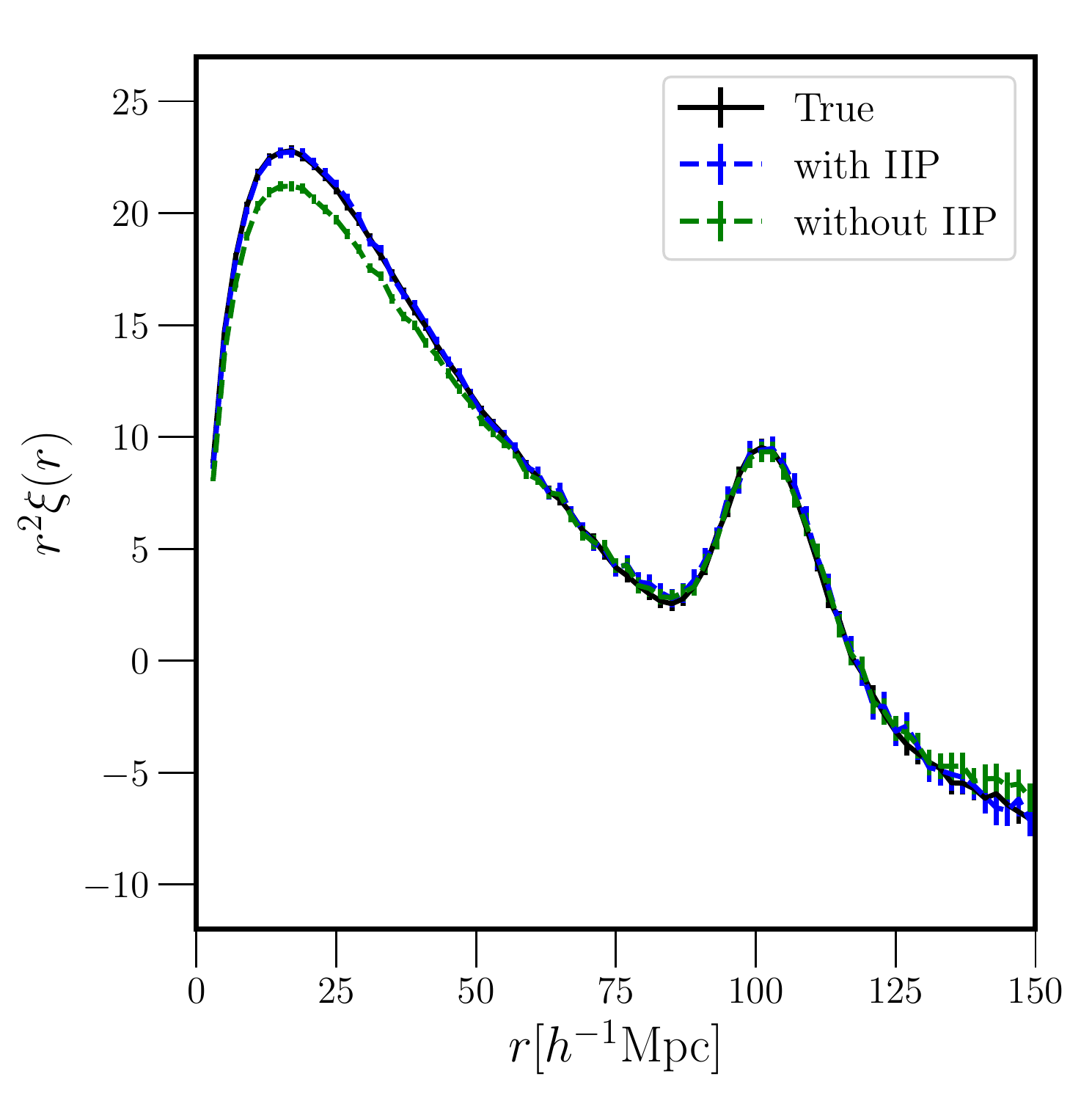}
    \includegraphics[width=0.48\textwidth]{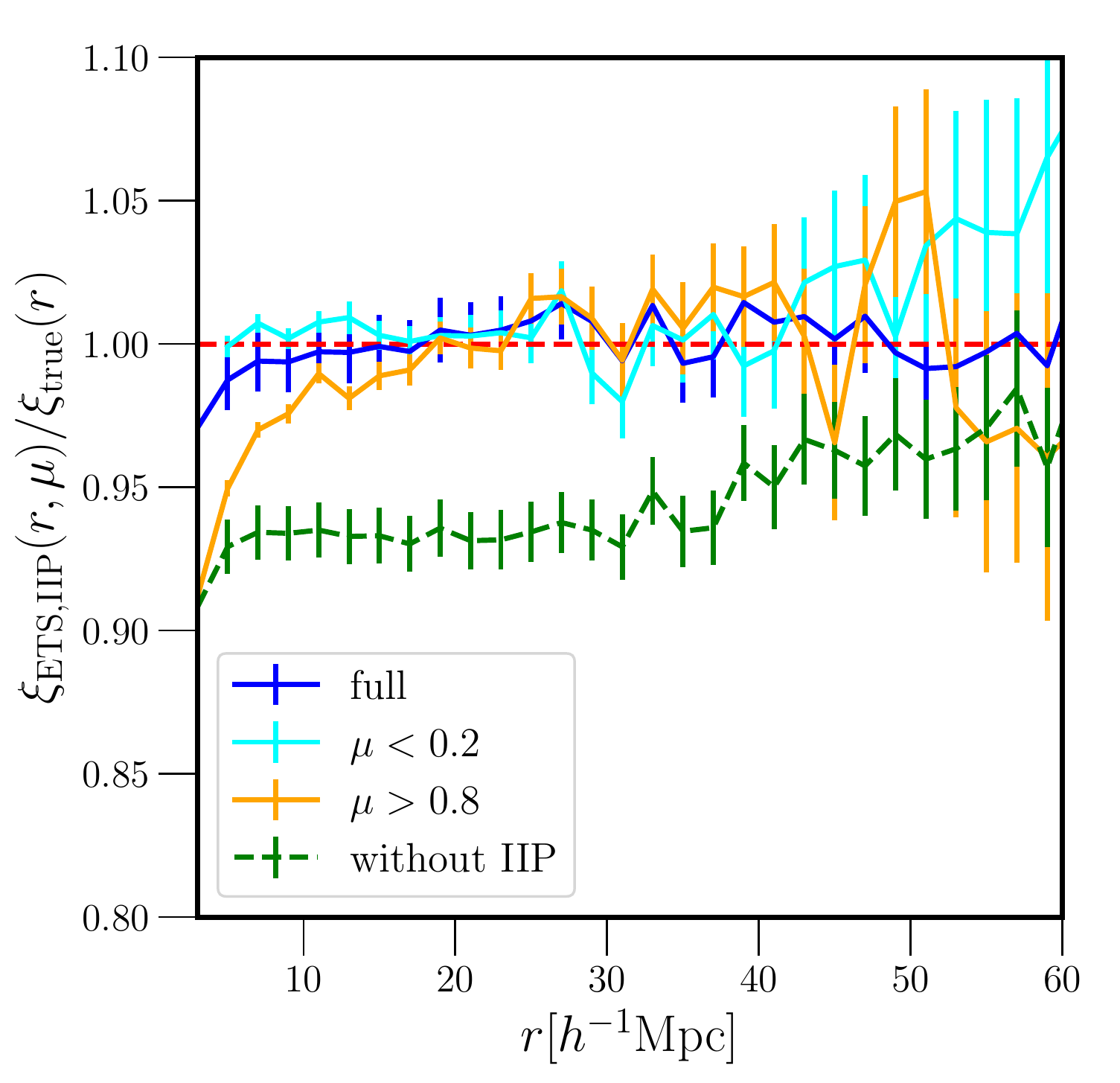}
    \caption{\label{fig:ets:iip} Same as Fig.~\ref{fig:uniform:iip}, but with the ETS. The probabilities are derived from running the ETS 100 times on all 100 mock catalogs.}
\end{figure*}

What is the probability that a target galaxy is observed by the ETS? Each target galaxy in the mock catalog has a probability to be observed. We calculate this probability by running the ETS 100 times on one mock catalog. 
Fig.~\ref{fig:ets:prob} shows the probabilities to be observed for all targets in the different redshift slices. 
The probabilities have a wide distribution as a result of the fiber assignments via the ETS, compared to the discrete probabilities, 1 or $2/n_{i,{\rm target}}$ ($n_{i,{\rm target}}\ge 3$), for the toy model of tiles and fibers in Fig.~\ref{fig:pic2}. 
We find that the probabilities are similar for all the redshift slices. 
The weight for each target is given by the inverse of these probabilities.

In Fig.~\ref{fig:ets:iip}, we show that the IIP method still recovers the underlying correlation functions from realistic mocks using the ETS to better than 1\% accuracy at all separations (except the smallest one) for the angle-averaged correlation functions. 
The transverse-direction ($\mu<0.2$) is also recovered well, whereas the LOS direction ($\mu>0.8$) is recovered only at $r>10~h^{-1}{\rm Mpc}$.

\subsection{Recovery of the redshift-space distortion effect}
\label{sec:final_result}
\begin{figure*}
\centering
    \includegraphics[width=0.4\textwidth]{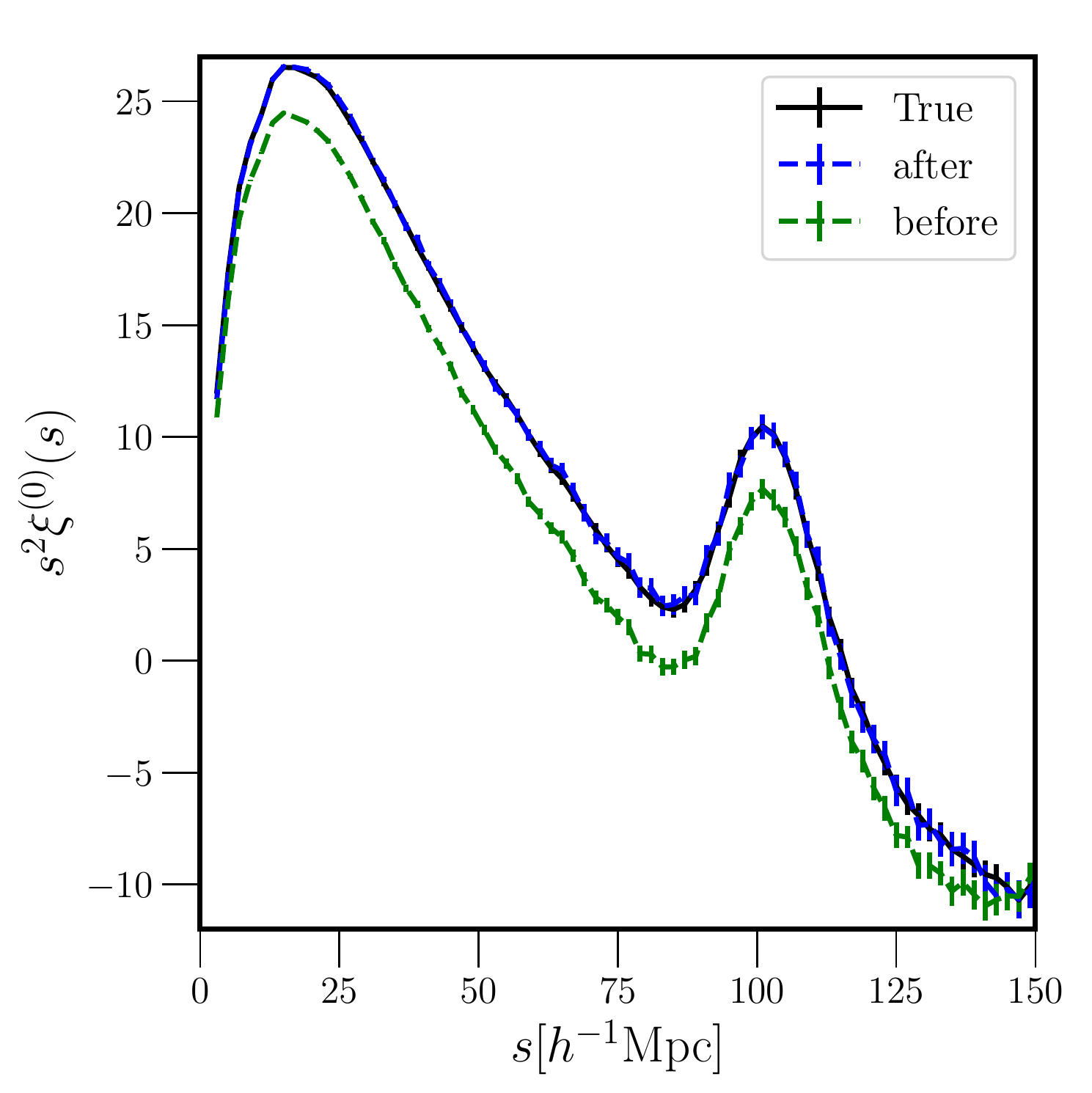}\includegraphics[width=0.4\textwidth]{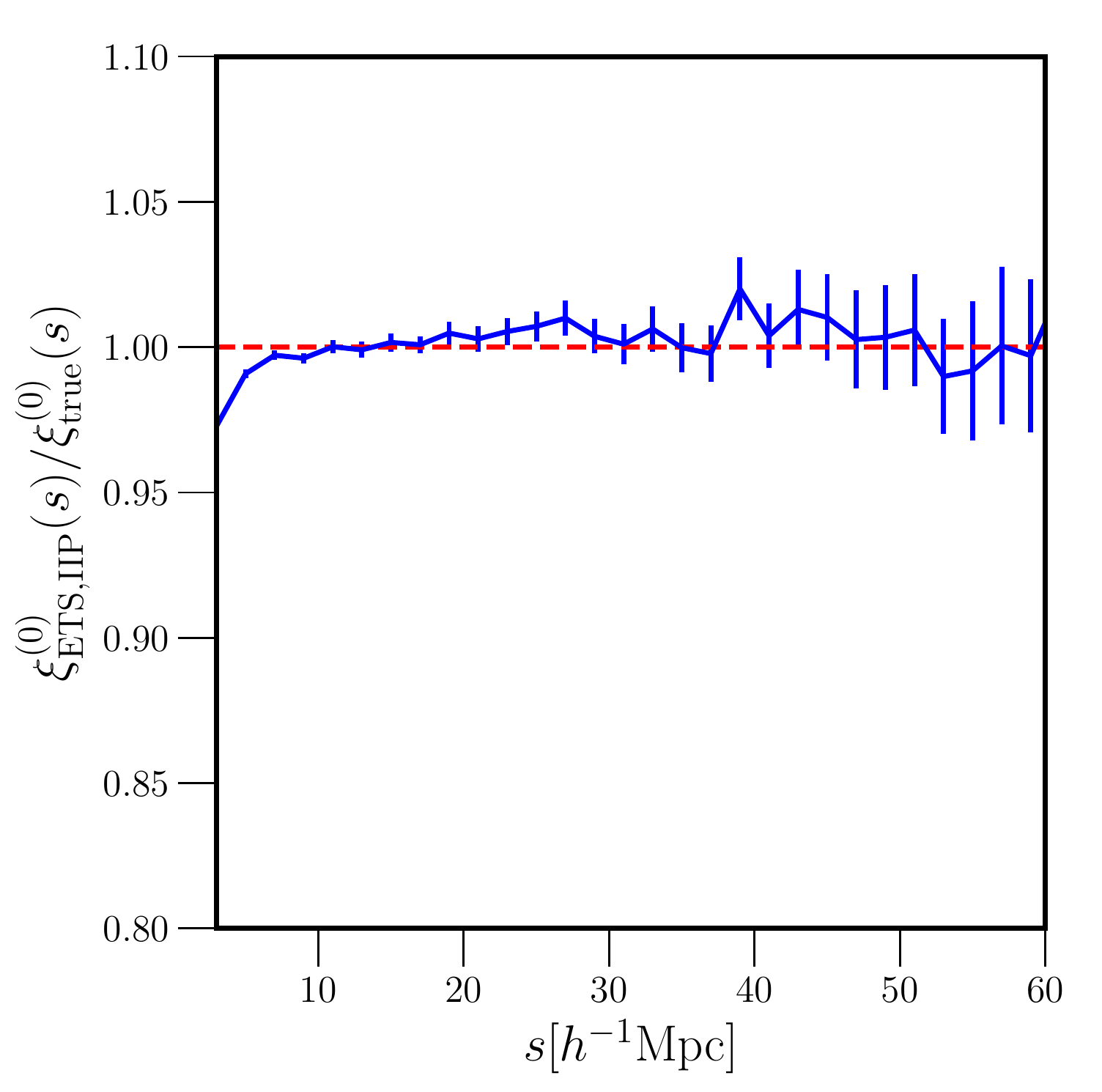}
    \includegraphics[width=0.4\textwidth]{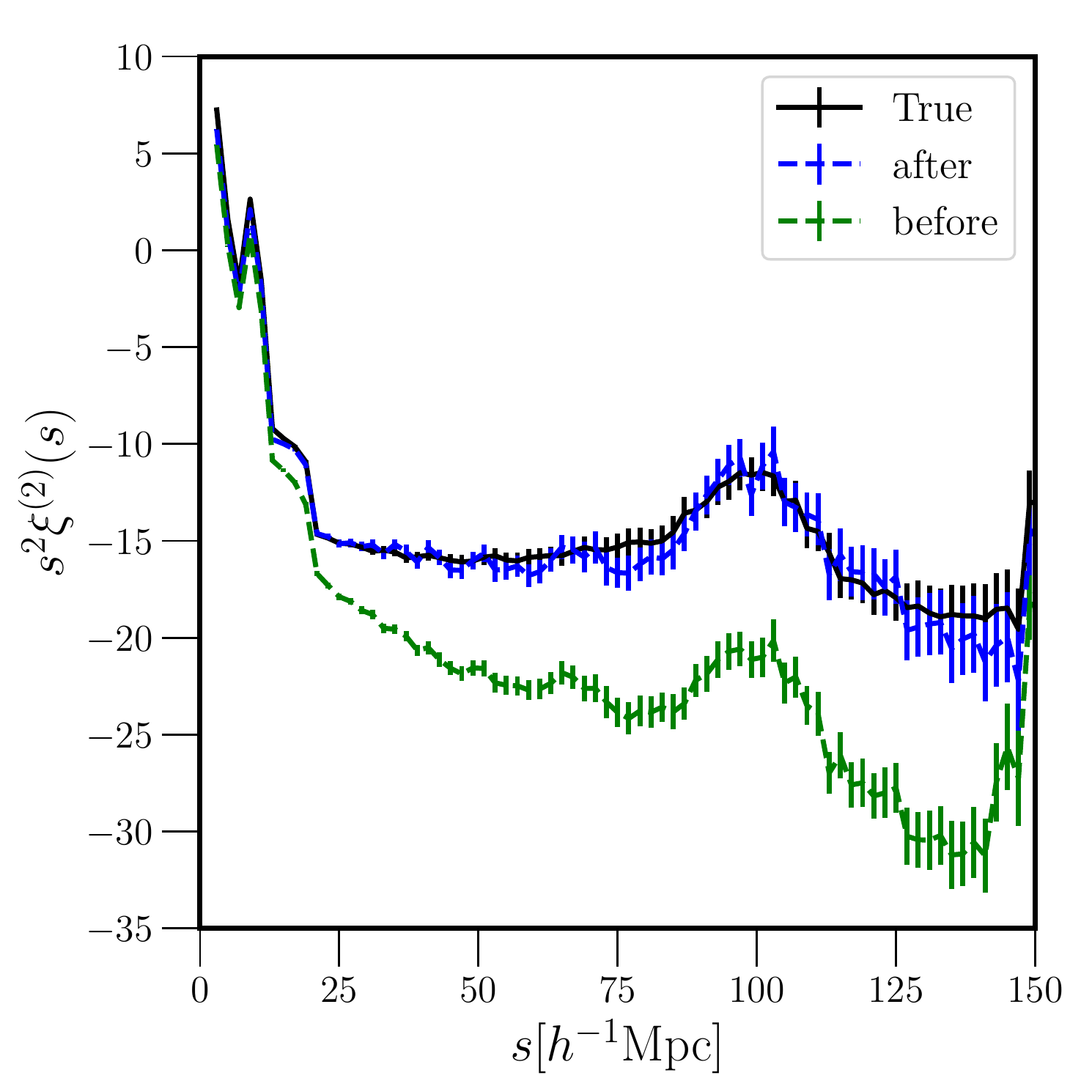}\includegraphics[width=0.4\textwidth]{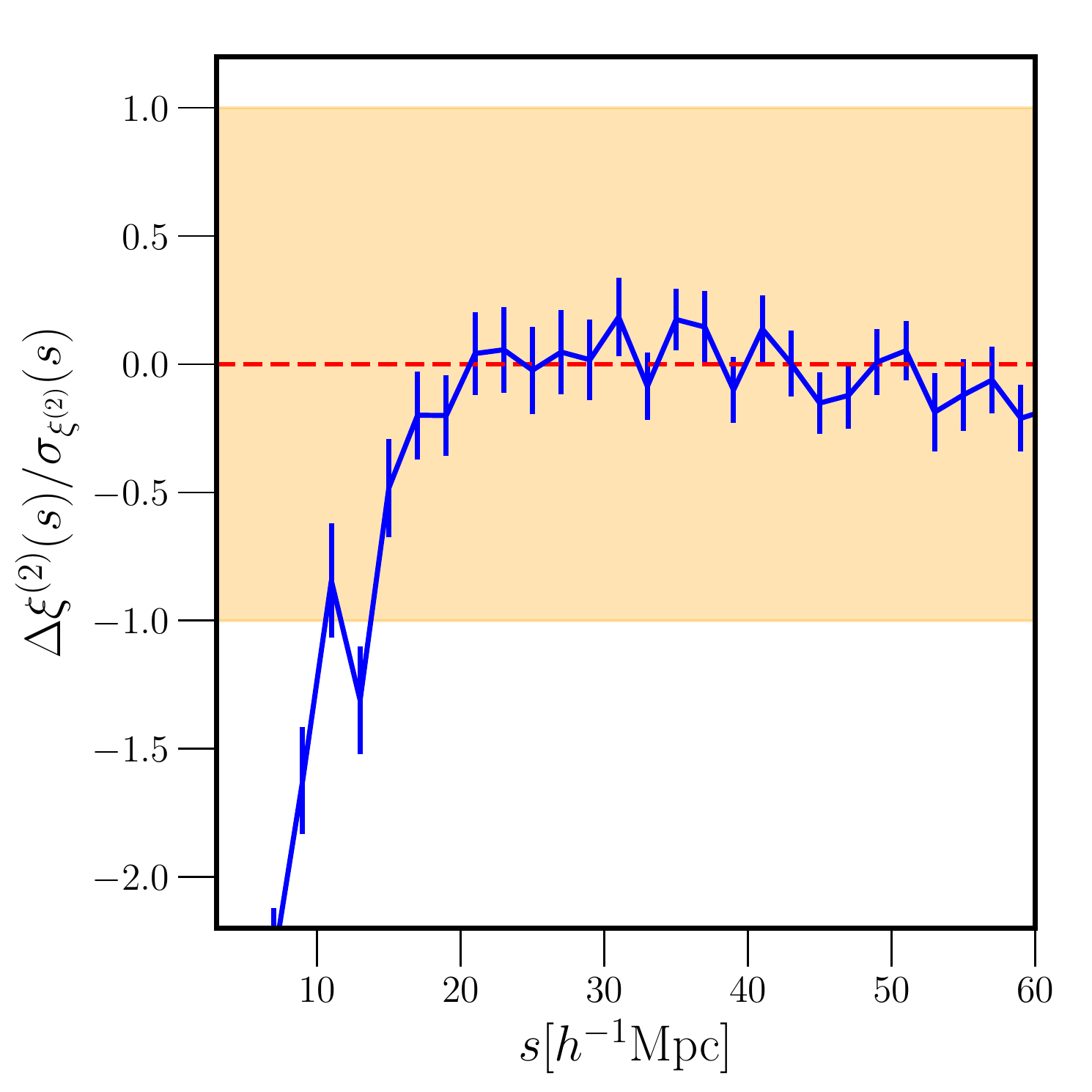}
    \caption{\label{fig:ets:iip_s} Monopole (top left) and quadrupole (bottom left) of correlation functions before (green) and after (blue) applying the weights to correct the tiling and fiber assignment effects.  The black data show the underlying correlation functions.
    The top right panel shows the ratios of the monopole correlation functions in the top left panel. The bottom right panel shows the differences of the quadrupole correlation functions in the bottom left panel, divided by the standard deviation of the 560~deg$^2$ survey area.
    We show the differences (rather than the ratios) for the quadrupole correlation functions because they have a zero-crossing in the amplitude at $10~h^{-1}{\rm Mpc}$.
    Here we show the results for the $z=1.3$ slice, but we find similar results for the other redshift slices. 
    }
\end{figure*}

We now introduce the peculiar velocity field to the mock catalogs and investigate the impact of tiling and fiber assignment on the redshift-space distortions.
Fig.~\ref{fig:ets:iip_s} shows that the weights to correct the tiling effect and the IIP recover the underlying monopole and quadrupole correlation functions to better than 1\% and 5\%
%
%
accuracy at $r>10~h^{-1}{\rm Mpc}$ respectively.
We also find that the same method corrects for the tiling and fiber assignment effects for all the other redshift slices with similar accuracy.

\section{Conclusion}
\label{sec:discussion}
In this paper, we have investigated the tiling and fiber-assignment effects on galaxy clustering. 
The density fluctuation of long-wavelength is lost due to the tiling effects, and the target galaxies in under-dense regions are preferentially observed due to fiber-assignment effects. 
Even though we analyze these problems with the PFS survey in mind, they are not specific to the PFS survey: All fiber-fed spectroscopic galaxy surveys that are deep and have low completeness will suffer from the same problems.
While the fiber collision problem in previous galaxy surveys such as BOSS alters the angular clustering, the tiling and fiber-assignment effects alter the clustering along the LOS direction more.

To mitigate the effects on clustering, we have applied two weighting methods. 
One is the weight by the number of target galaxies in each tile, which reconstructs the long-wavelength density fluctuations beyond the size of the FoV.
The other is the weight by the inverse of the probability that a target galaxy is observed by the fiber assignment algorithm (the IIP method \cite{Smith_etal2019}). The IIP method down-weights observed galaxies in under-dense regions.
We have demonstrated that these two weighting methods unbias both the monopole and quadrupole correlation functions to better than 1\% and 5\% accuracy at $r>10~h^{-1}{\rm Mpc}$ respectively.

\acknowledgments
We thank the PFS Collaboration, especially the members of the Cosmology Science Working Group, for useful discussion and feedback on this project. This work was supported in part by World Premier International Research Center Initiative (WPI Initiative), MEXT, Japan, JSPS KAKENHI Grant Numbers JP15H03654, JP15H05887, JP15H05893, JP15H05896, JP15K21733, JP16K17659, JP17K14273, JP18K13578, and JP19H00677, and Japan Science and Technology Agency CREST JPMHCR1414.


\bibliographystyle{JHEP}
\bibliography{refs}
\end{document}